 \documentclass[preprint2]{aastex}

\slugcomment{Submitted to ApJ}

\shorttitle{No Doppler-activity correlations on Kapteyn's star}
\shortauthors{Anglada-Escude, Tuomi, et al.}

\begin{document}


\title{No evidence for activity correlations in the
radial velocities of Kapteyn's star}


\author{
G. Anglada-Escud\'e\altaffilmark{1,2}, 
M. Tuomi \altaffilmark{1},
P. Arriagada\altaffilmark{3},
M. Zechmeister\altaffilmark{2},
J. S. Jenkins\altaffilmark{4},
A. Ofir\altaffilmark{5},
S. Dreizler\altaffilmark{6},
E. Gerlach\altaffilmark{7},
C.~J. Marvin\altaffilmark{6},
A. Reiners\altaffilmark{6},
S.~V. Jeffers\altaffilmark{6},
R. Paul Butler\altaffilmark{3},
S.~S. Vogt\altaffilmark{8},
P.~J. Amado\altaffilmark{9},
C. Rodr\'iguez-L\'opez\altaffilmark{9},
Z. M. Berdi\~nas\altaffilmark{9},
J. Morin\altaffilmark{10},
J.~D. Crane\altaffilmark{11},
S.~A. Shectman\altaffilmark{11},
%I.~B. Thompson\altaffilmark{11},
M. D\'iaz\altaffilmark{4},
%E. Rivera\altaffilmark{8},
L.~F.~Sarmiento\altaffilmark{6},
H.~R.~A. Jones\altaffilmark{1}
%J. Teske\altaffilmark{3}
}


\altaffiltext{1}{
Centre for Astrophysics Research,
University of Hertfordshire,
College Lane, AL10 9AB, Hatfield, UK}
\altaffiltext{2}{
School of Physics and Astronomy, Queen Mary, University of London,
327 Mile End Rd. London, United Kingdom
}
\altaffiltext{3}{
Carnegie Institution of Washington,
Dept. of Terrestrial Magnetism,
5241 Broad Branch Rd. NW, 20015,
Washington D.C., USA}
\altaffiltext{4}{
Departamento de Astronom\'ia, Universidad de Chile,
Camino El Observatorio 1515, Las Condes, Santiago, Chile,
Casilla 36-D.}
\altaffiltext{5}{
Department of Earth and Planetary Sciences, Weizmann Institute of Science, 234 Herzl St., Rehovot 76100, Israel
}
\altaffiltext{6}{
Universit\"{a}t G\"{o}ttingen,
Institut f\"ur Astrophysik,
Friedrich-Hund-Platz 1,
37077 G\"{o}ttingen, Germany
}
\altaffiltext{7}{
Institut f\"ur Planetare Geod\"asie
Technische Universit\"at Dresden
01062, Dresden, Germany}
\altaffiltext{8}{
UCO/Lick Observatory, University of California,
Santa Cruz, CA, 95064, USA}
\altaffiltext{9}{
Instituto de Astrof\'isica de Andaluc\'ia-CSIC,
Glorieta de la astronom\'ia S/N, 18008, Granada, Spain}
\altaffiltext{10}{
LUPM-UMR5299, CNRS \& Universit\' de Montpellier,
Place E. Bataillon, Montpellier, F-34095, France
}
\altaffiltext{11}{
Carnegie Institution of Washington, The Observatories, 
813 Santa Barbara Street, Pasadena, CA 91101-1292, USA
}


\begin{abstract} 

Stellar activity may induce Doppler variability at the level of a few m/s which
can then be confused by the Doppler signal of an exoplanet orbiting the star. To
first order, linear correlations between radial velocity measurements and
activity indices have been proposed to account for any such correlation. The
likely presence of two super-Earths orbiting Kapteyn's star was reported in
\citet{anglada:2014a}, but this claim was recently challenged by
\citet{robertson:2015b} arguing evidence of a rotation period (143 days) at
three times the orbital period of one of the proposed planets (Kapteyn's b,
P=48.6 days), and the existence of strong linear correlations between its
Doppler signal and activity data. By re-analyzing the data using global
optimization methods and model comparison, we show that such claim is incorrect
given that; 1) the choice of a rotation period at 143 days is unjustified, and
2) the presence of linear correlations is not supported by the data. We conclude
that the radial velocity signals of Kapteyn's star remain more simply explained
by the presence of two super-Earth candidates orbiting it. We also advocate for
the use of global optimization procedures and objective arguments, instead
of claims lacking of a minimal statistical support. 

\end{abstract}


\section{Introduction}

Recently, the search for low-amplitude signals in radial velocity time-series
has reached the point where detection of Doppler signals at the level of 1m/s or
less is technically possible \citep{pepe:2011, tuomi:2013}.  Along with this
rise in precision have come claims, and counter-claims, of the detection of
planetary systems containing very low-mass  planets (e.g. $\alpha$~Centauri,
\citealp{dumusque:2012}, \citealp{hatzes:2013};  HD~41248
\citealp{jenkins:2013b,jenkins:2014}, \citealp{santos:2014}; GJ~581
\citealp{mayor:gj581:2009}, \citealp{robertson:2014a}, \citealp{anglada:2015a}).
Given the sensitive nature of these works, it is clear more work must be done 
to develop a clear structure for what constitutes a Doppler signal detection 
and what does not.

It is known that stellar activity might induce spurious signals in precision
Doppler measurements \citep[eg.]{queloz:2001}. In particular, variability in
chromospheric activity indices are supposed to originate from localized active
regions on stars. Changes in the local properties of the visible surface of
stars can induce apparent Doppler shifts that do not necessarily average out
over time, producing apparent signals that might be mistaken as planets
\citep[eg.][]{hatzes:2002,bonfils:2007}. Theoretical and numerical simulations
suggest that variability on some of these indices should linearly correlate with
apparent radial velocity shifts \citep{boisse:2011,dumusque:2014}.
\citet{robertson:2014a} exploited this expected linear correlation to propose
that the planet candidate GJ~581d was caused by stellar variability by showing
some correlations of activity indices with residual time-series (all other
signals removed). Since residual time-series are not representative of the
original data, such conclusions were challenged by \citet{anglada:2015a}. In
response, \citet{robertson:2015a} admitted inconsistencies in their statistical
analysis but claimed that their interpretation of the data was physically more
sound. Along these lines, in \citet{robertson:2014b} and 
\citet{robertson:2015b} (RM15 hereafter) similar qualitative arguments were
provided to argue that several super-Earth mass planet candidates orbiting
nearby M-dwarf stars were likely to be spurious. In this paper we show that the
claims in \citet{robertson:2015b} are unsupported by a global fit to the data,
so such results should be regarded as inconclusive.

The data used in this paper comes directly from RM15 to replicate their setup as
closely as possible. The datasets in RM15 contain measurements obtained with the
HARPS and the HIRES spectrometers. These are different from the ones in
\citet{anglada:2014a} in the sense that RM15 includes additional spectroscopic
indices and, additionally, three HARPS epochs (out of 95) were removed. We also
include the analysis of V magnitude historical photometric measurements obtained
by the ASAS project \citep{asas}. A more detailed description of the
measurements are given in both papers and references therein.  We start by
reviewing possible periodic signals in the activity indices presented by RM15 in
Section \ref{sec:activity}. Section \ref{sec:model} introduces a minimal Doppler
model to include linear correlation terms caused by activity. To remove 
ambiguities about the framework used, we perform the analyses in a frequentist
(Section \ref{sec:likelihood}) and a Bayesian framework (Section
\ref{sec:bayesian}); both providing a consistent picture of no correlations in
either case. Section \ref{sec:nocorrelations} discusses the discrepancy between
our results and the analysis presented in RM15. A summary and concluding remarks
are given in Section \ref{sec:conclusions}.

\section{Possible signals in activity indices and ASAS photometry}
\label{sec:activity}

\begin{figure}[]
\center
\includegraphics[width=0.45\textwidth, clip]{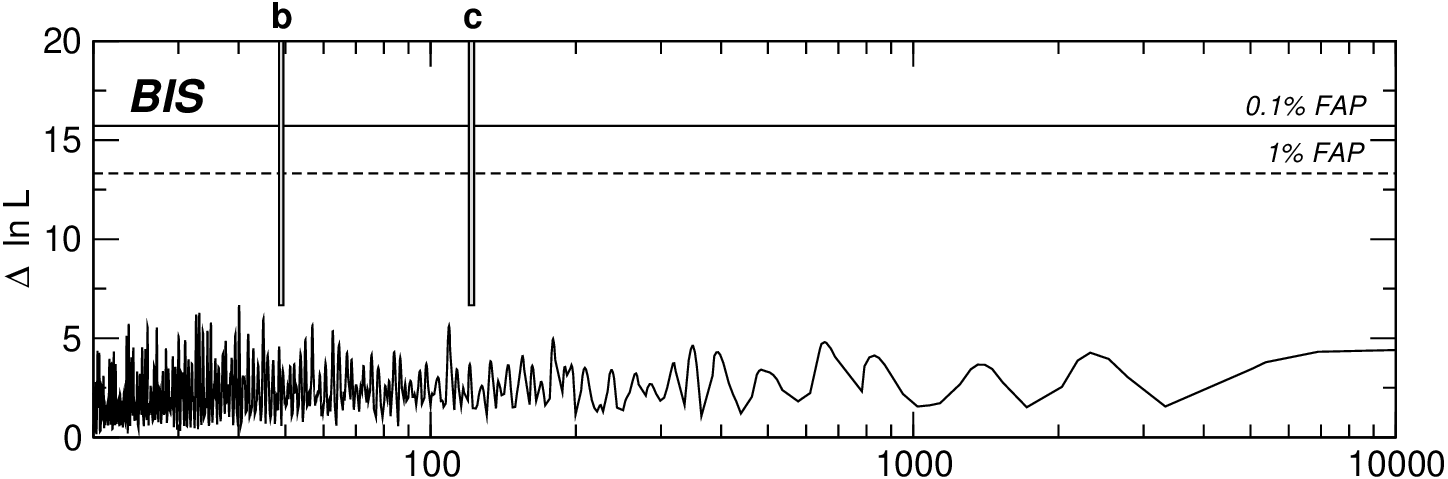}
\includegraphics[width=0.45\textwidth, clip]{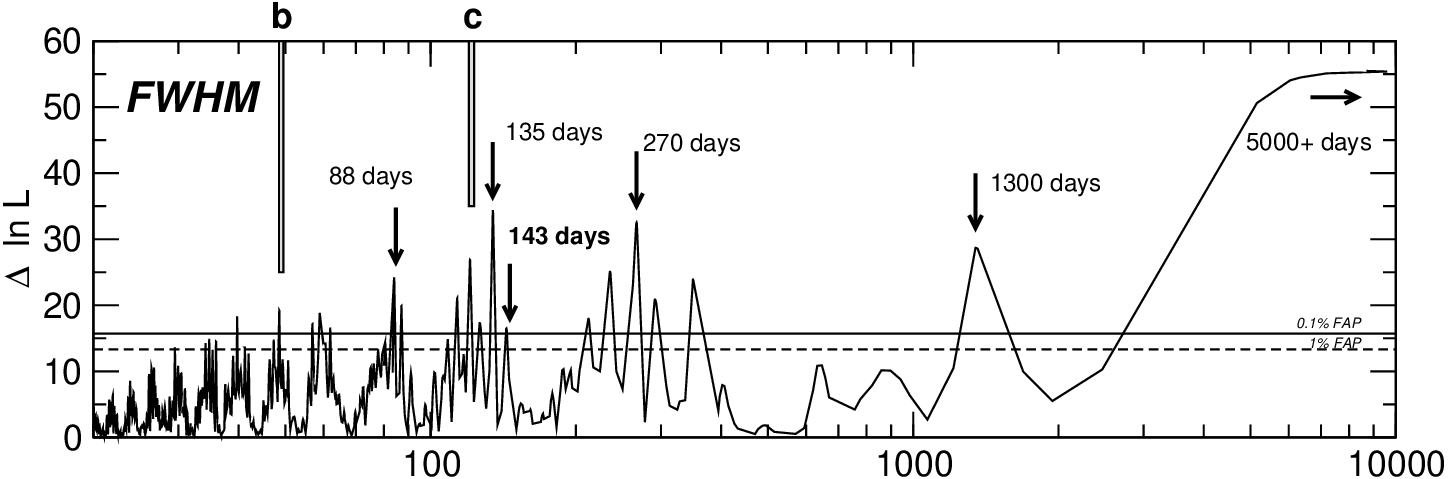}
\includegraphics[width=0.45\textwidth, clip]{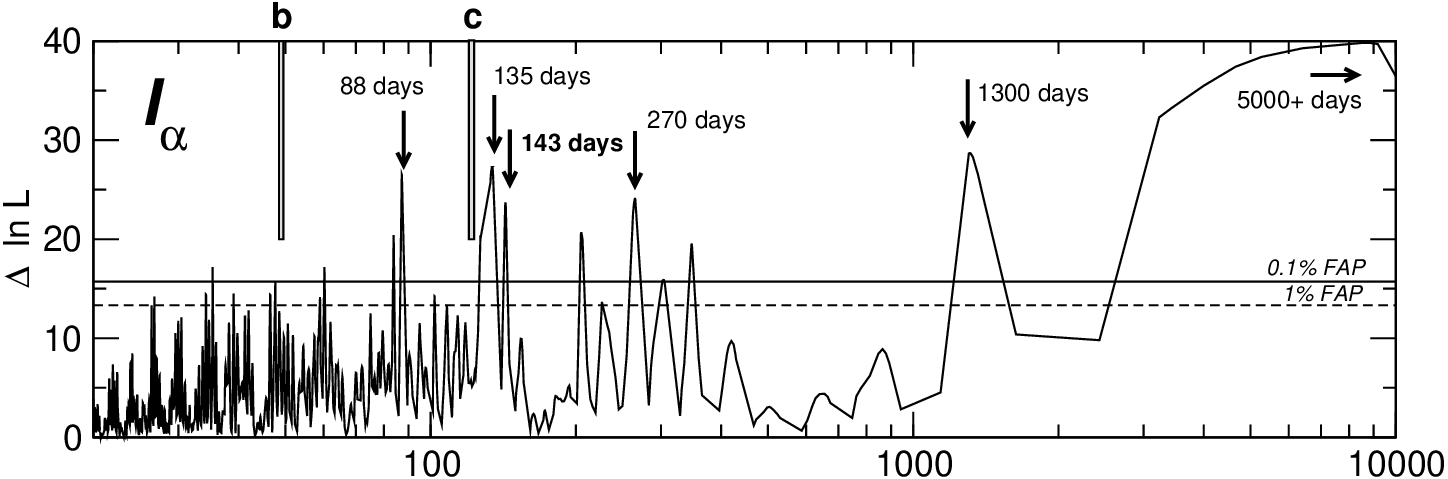}
\includegraphics[width=0.45\textwidth, clip]{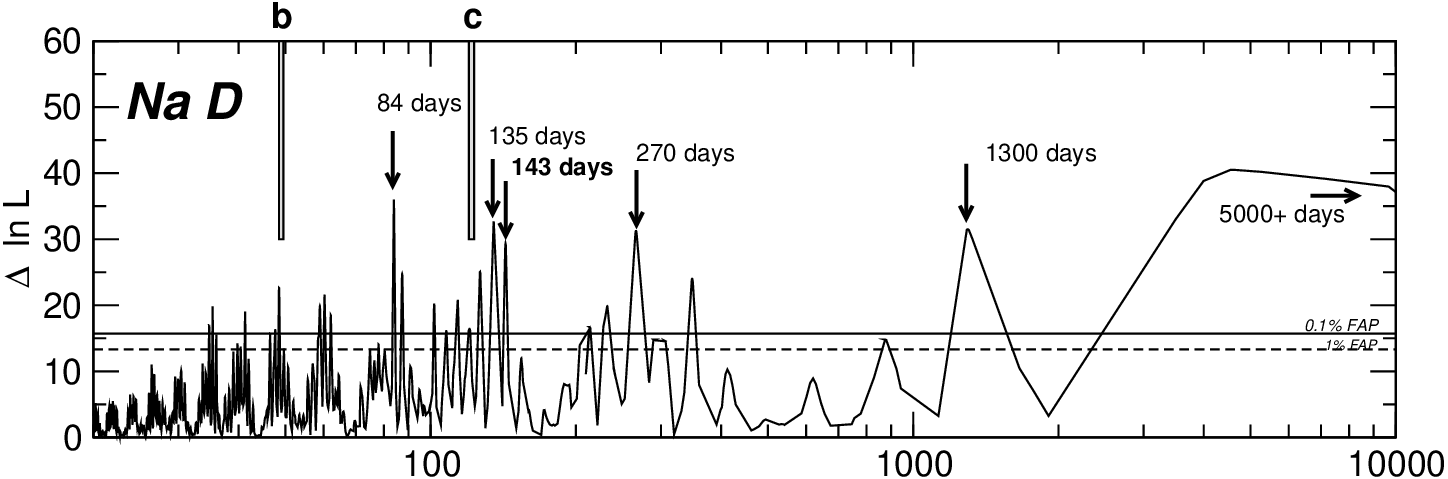}
\includegraphics[width=0.45\textwidth, clip]{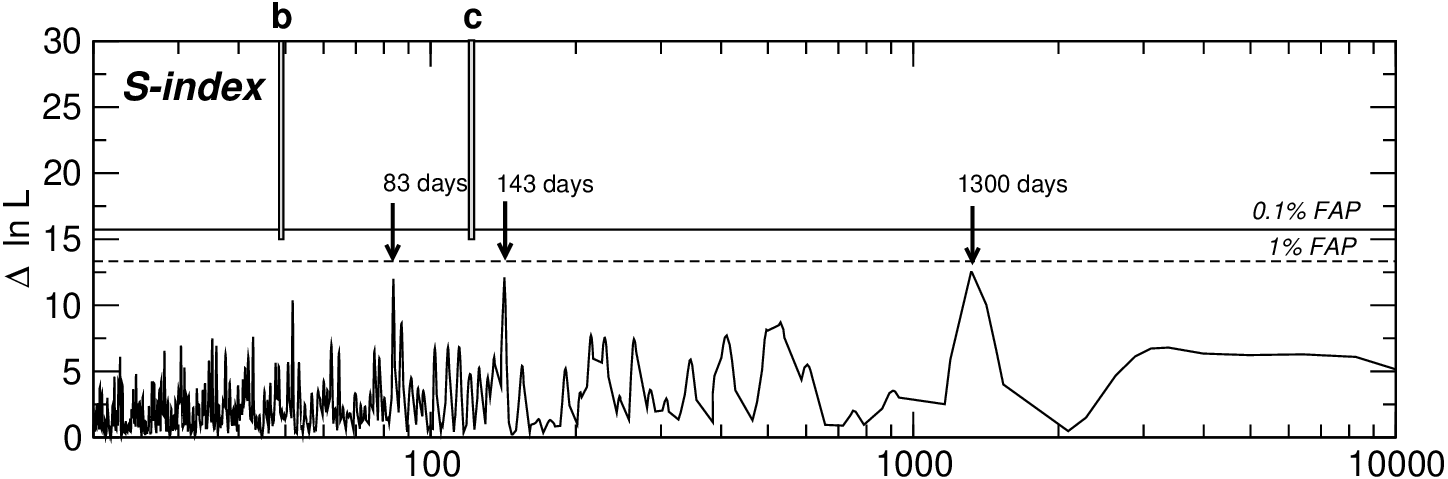}
\includegraphics[width=0.45\textwidth, clip]{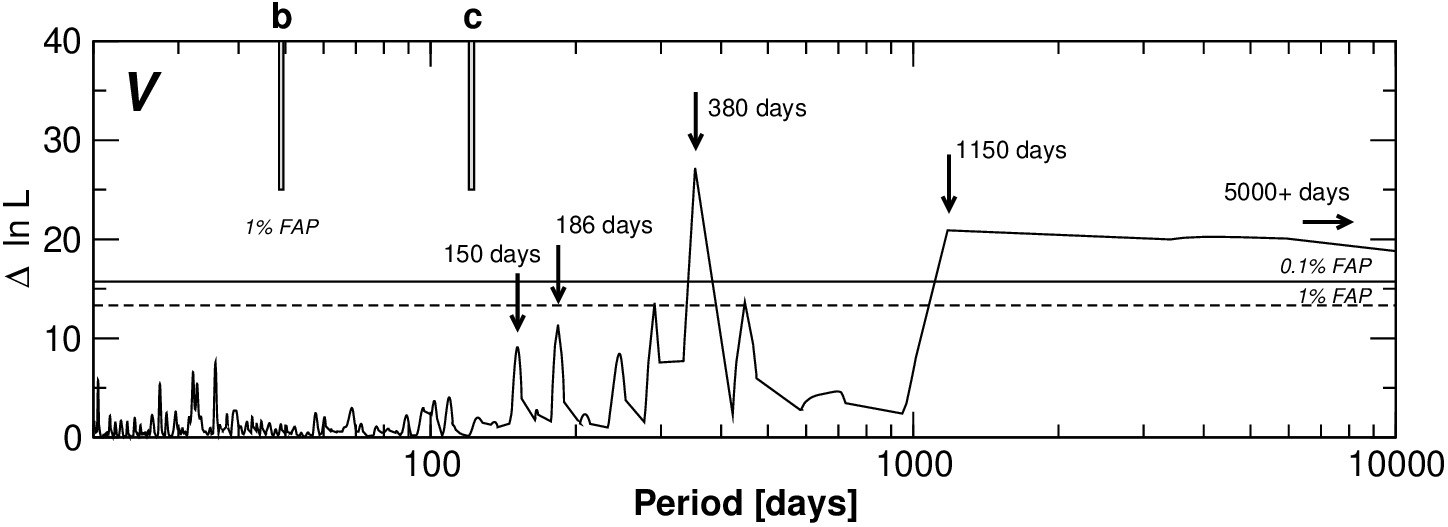}

\caption{Likelihood periodograms of the activity indices in
RM15 (from top to bottom; BIS, FWHM, $I_\alpha$, Na D,
S-index) and ASAS V band photometry (bottom). With the
exception of BIS, variability above 1\% FAP threshold
(horizontal line) is detected in all indices. Most relevant
possible periods in each activity index are flagged with
arrows. The similarities between periodograms in different
indices (long period trend, and possible signals between 80
and 300 days) suggest similar, non-strictly periodic stellar
variability in these time-ranges but does not point out to a
clearly preferred signal. }\label{fig:actper}

\end{figure}

We perform a likelihood periodogram analysis of the activity indices as provided
by RM15 to verify the claim of a \textit{clear} rotation period at 143 days.
Likelihood ratio periodograms solves for all the free parameters of the model at
the same time when a signal is injected over a list of trial periods (x-axis).
Such periodograms are a generalization of Lomb-Scargle periodograms
\citep{scargle:1982} to account for models more complex than a single sinusoid
\citep{baluev:2009}, including parameters of the noise model (eg. extra white
noise for the activity data). The signal producing the highest improvement of
the maximum log-likelihood statistic (y-axis) would be the preferred one and its
significance can then be assessed using the recipes introduced by
\citet{baluev:2009,baluev:2013}, producing  analytic estimates of the false
alarm probability of detection (or FAP). As a general rule, signals above a FAP
threshold of 1 \% can be considered significant, but a more conservative
threshold of 0.1\% is sometimes used. We present both in all the periodograms
presented throughout the paper. In the case of activity data, we assume that the
signal is modelled by: one constant (equivalent to the mean of the time-series),
one sinusoid (phase and amplitude are free parameters), and an extra white noise
parameters added in quadrature to the nominal uncertainties of each measurement.
As mentioned by RM15, nights with several measurements might be overweighted and
bias the signal searches. To account for this, we present the analysis using
night averages only (45 independent epochs). Our conclusions however didn't
differ substantially if all datapoints were included.

The activity indices provided in RM15 include BIS, FWHM, I$_\alpha$, Na D,
S-index. The first two are measurements of the shape of the mean spectral line
(BIS and FWHM represent asymmetry and width respectively), which can potentially
trace activity-induced features on the stellar photosphere. The last three ones are
measurements of the chromospheric emission of the star at the H$_\alpha$
(I$_\alpha$),  Sodium D$_1$ and D$_2$ lines (Na D), and Calcium H+K lines
(S-index). Chromospheric indices are also supposed to trace the presence of
active regions on the star that might be responsible for apparent Doppler
shifts. More precise definitions and possible connection to activity-induced
signals are given in RM15 and references therein.  The results of signal
searches on the five indices used by RM15 (plus available V band photometry from
the ASAS survey) are summarized in Figure \ref{fig:actper}. 

No significant periodicity is detectable in BIS. Several other indices show
multiple peaks above the 1\% and 0.1\% FAP thresholds (horizontal dashed and
solid lines, respectively). However, several of the peaks have similar $\Delta$
ln-L values, meaning that they satisfy the data similarly well. The only
exception is the long period trend (marked as 5000+ days in
Fig~\ref{fig:actper}), which in some cases produces a much larger improvement of
the likelihood (eg. FWHM and $I_\alpha$; second and third panels from the top,
respectively). Although the periodograms in RM15 also show a likely long period
trend in several indices, this evidence was disregarded as irrelevant in RM15 by
using generalistic arguments that are not supported by the literature. That is,
most stars in the M-dwarf sub-sample of the HARPS-GTO program (Kapteyn's star is
part of it) were found to show chromospheric variability in similar indices over
long time-scales by \citet{dasilva:2012}.

In summary, signals at 5000+, 1100, 270, 135 and 88 days would explain the
activity data equally well (even better depending on the index). Given this
ambiguity the preferred periods in the various activity indices, the choice made
by RM15 for a rotation period at 143 days seems rather arbitrary.

\section{Search for correlations in the Doppler data}
\subsection{Model}\label{sec:model}

The next step in RM15's analysis was to assess the significances of linear
correlations of the Doppler signals with the activity indices. We implement
linear correlations by adding a linear relationship between the radial
velocities and activity data by using the following model 

\begin{eqnarray}
v(t) = M(\vec{\theta}, t) + \sum_i c_i\,I_i ,
\label{eq:model} \end{eqnarray} 

\noindent where $M$ contains all the Doppler variability
modeled by Keplerian signals, and $\vec{\theta}$ lists the
usual parameters used in RV modelling \citep[see][as an
example]{tuomi:2013}. Activity measurements obtained
simultaneous to $v(t)$ are I$_i$, where $i$ is added over all
the activity indices under consideration. As discussed before,
these indices include i=BIS, FWHM, I$_\alpha$, Na D, S-index. 

Given a model, one can search for the combination of parameters
that optimize a figure of merit (global optimization), and then
decide whether the inclusion of a correlation term or a planet
is warranted given the improvement of the reference statistic.
As long as global optimization is applied (all parameters
adjusted simultaneously), there are various ways to assess
significance of planetary signals or correlations using either
\textit{Bayesian} or \textit{frequentist} approaches
\citep{anglada:2012c}. A Bayesian approach consists of
assessing which model has the highest probability given the
data. Frequentist confidence tests evaluate the chances of
obtaining an improvement of a statistic by an unfortunate
combination of random errors. While RM15 show some apparent
correlations when representing one Doppler signal against some
of their activity data, the significance of those correlations
was never established using model comparison. The next two sections
show that the correlations claimed in RM15 are not significant
when a global fit to the data is obtained in either framework.

\subsection{Frequentist analysis}
\label{sec:likelihood}

\begin{figure}[htb]
\centering
\includegraphics[width=0.45\textwidth, clip]{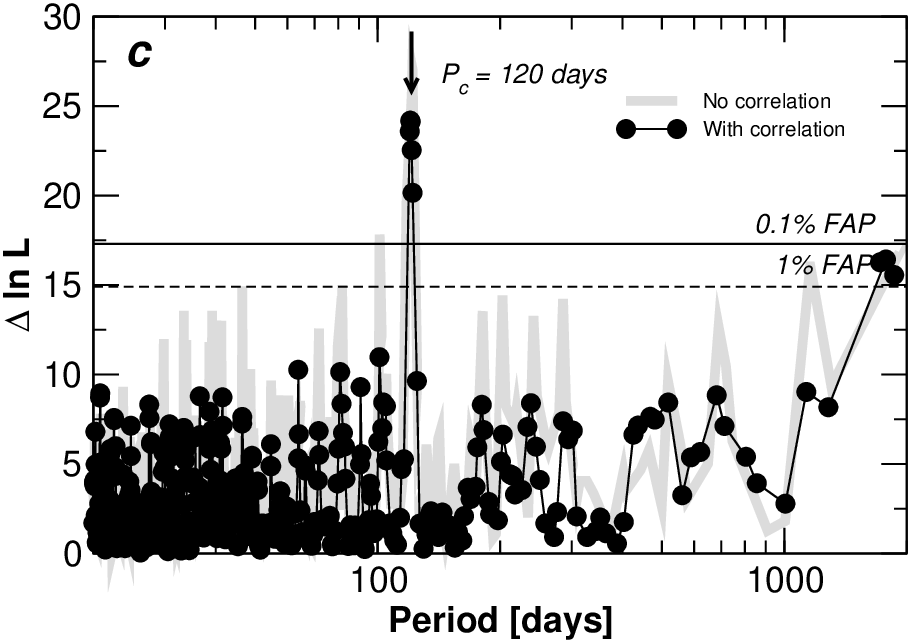}
\includegraphics[width=0.45\textwidth, clip]{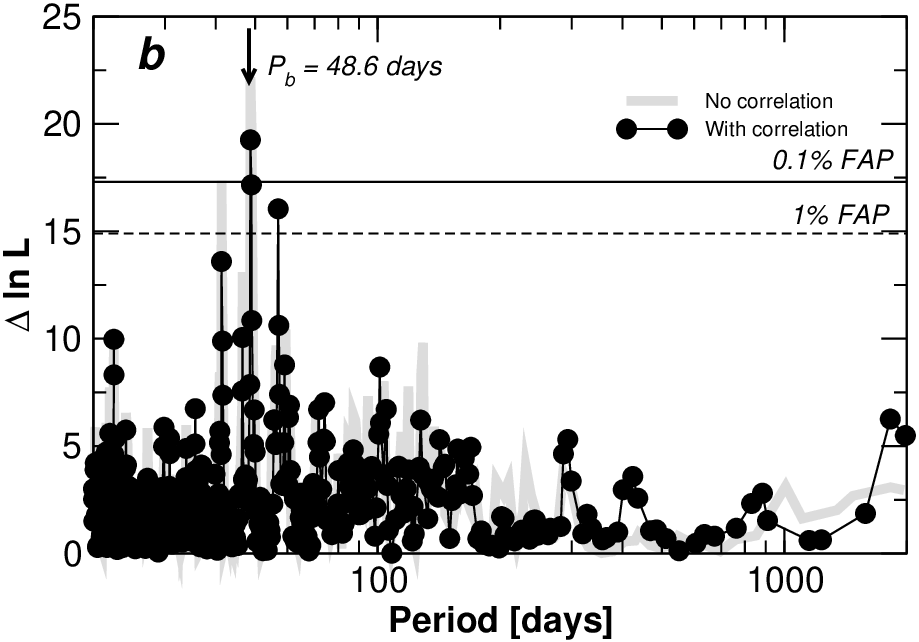}

\caption{Likelihood-ratio periodograms for first (top,
Kapteyn's c, k=1 planet) and second Doppler signals (bottom, Kapteyn's
b, k=2 planets), without linear correlations (gray) and
including linear correlations with the $I_{\alpha}$ index
(connected black dots). The peaks for the Doppler signals
remain above the 1\% and 0.1\% FAP thresholds in both cases.}
\label{fig:periodograms} 

\end{figure}

In RM15, the strongest apparent correlation was reported to be in the
chromospheric flux as measured by their $I_\alpha$ index. In
Fig.~\ref{fig:periodograms} we present likelihood ratio periodograms of the
combined HARPS and HIRES data (each data-set has its own linear correlation
coefficient as a free parameter). As shown in Fig.~\ref{fig:periodograms}, the
significance of both signals (120 and 48.6 days) remain well above the 0.1\% FAP
threshold, even when linear correlations are included in the model.
If linear correlations could explain the data better, adding a Keplerian signal
would not improve the fit substantially and its peak would be suppressed below
threshold. A similar result is obtained by using the other activity indices from
RM15 (omited here for brevity). In summary, the likelihood analysis indicates
that the linear correlation model cannot account for the presence either Doppler
signals.

\subsection{Bayesian analysis}
\label{sec:bayesian}

\begin{figure*}[htb]
\includegraphics[angle=270, width=0.195\textwidth]{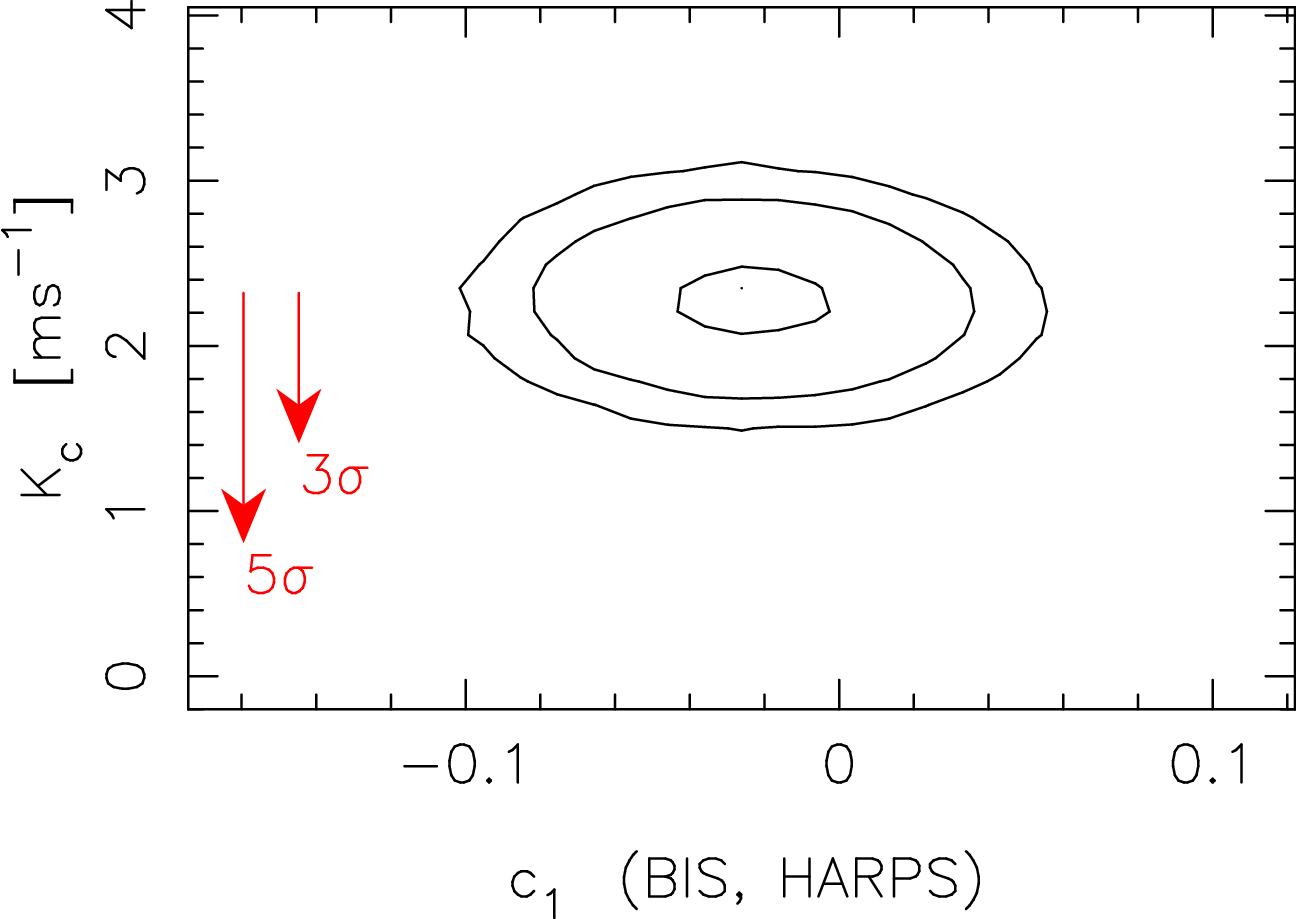}
\includegraphics[angle=270, width=0.195\textwidth]{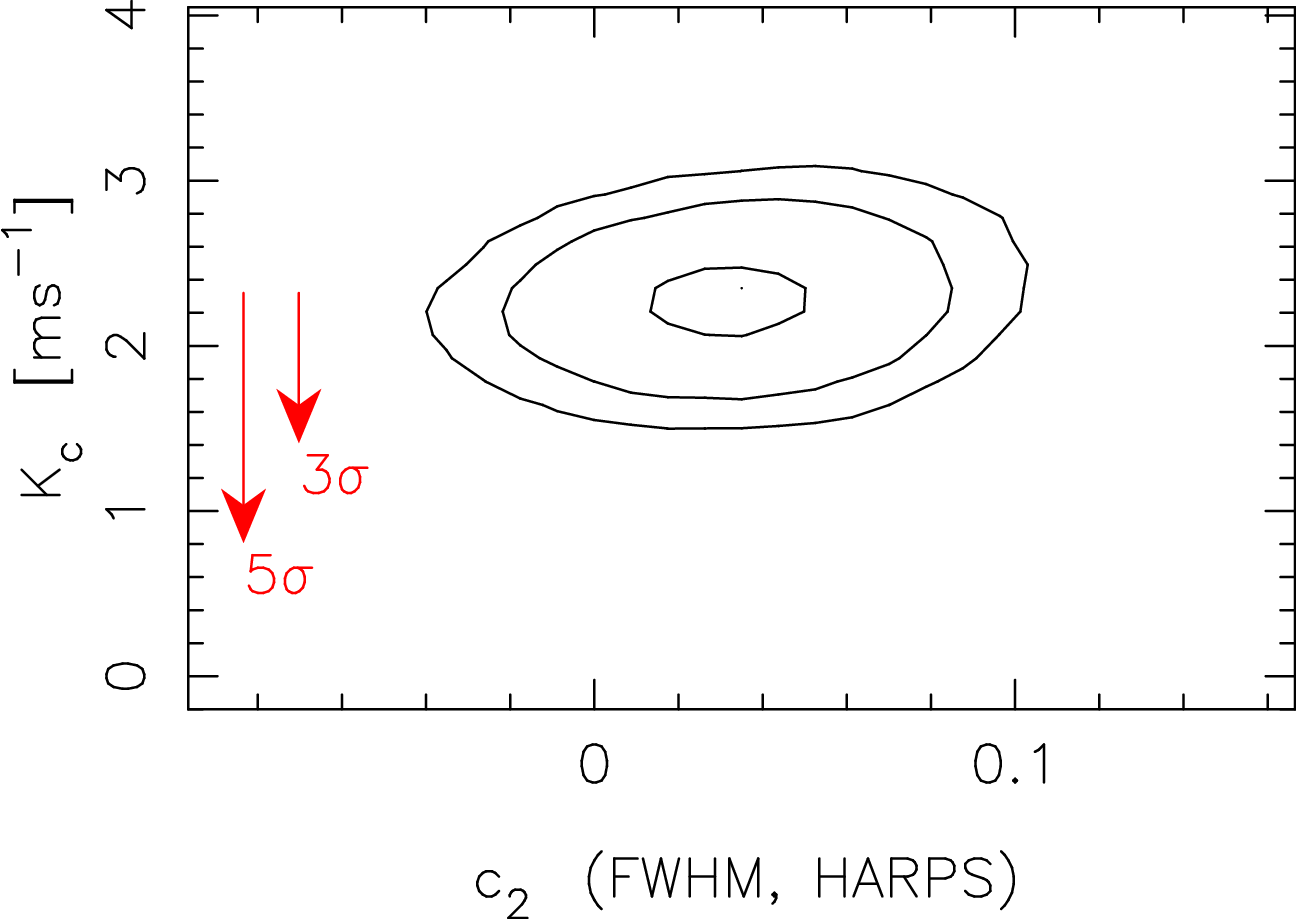}
\includegraphics[angle=270, width=0.195\textwidth]{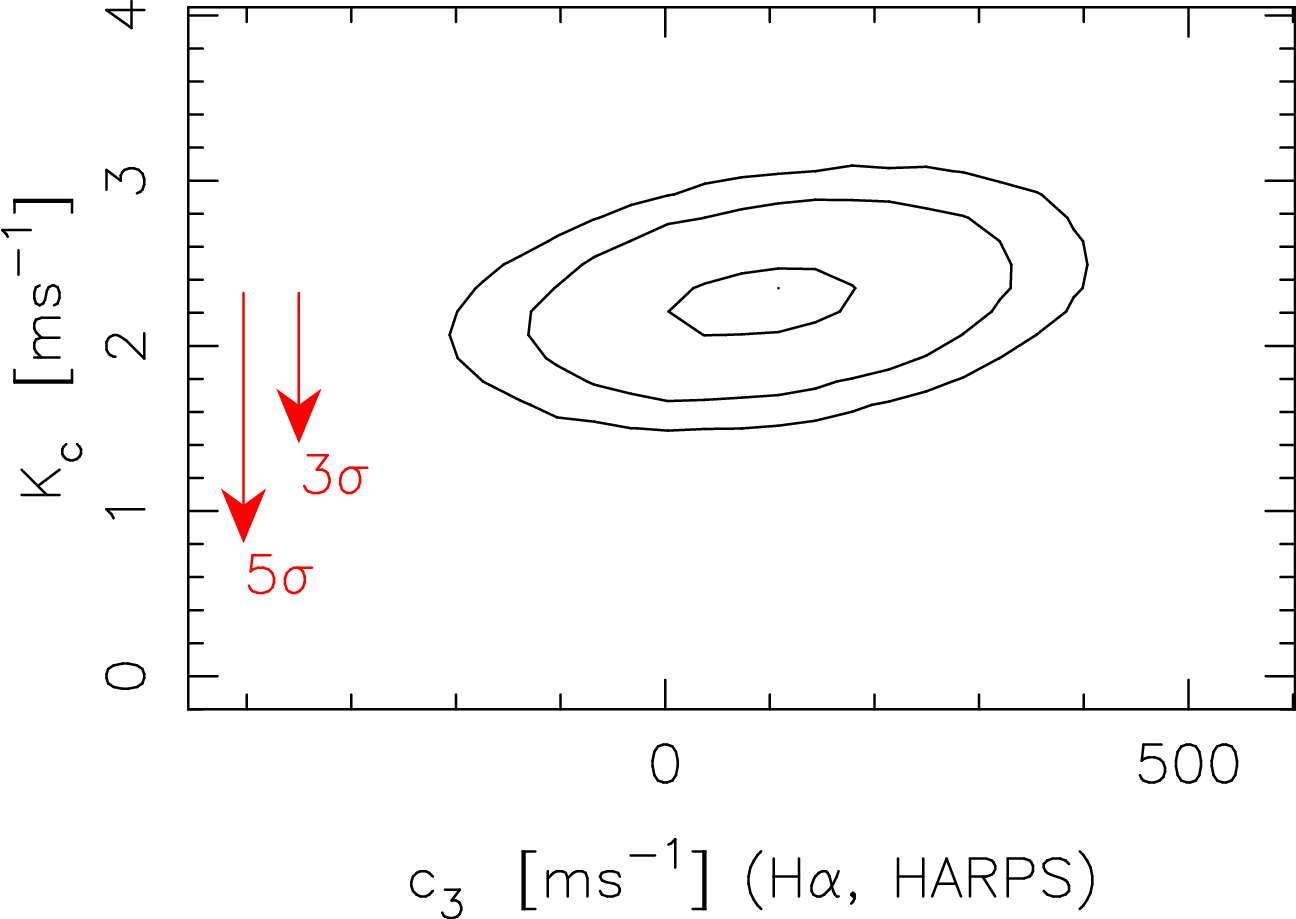}
\includegraphics[angle=270, width=0.195\textwidth]{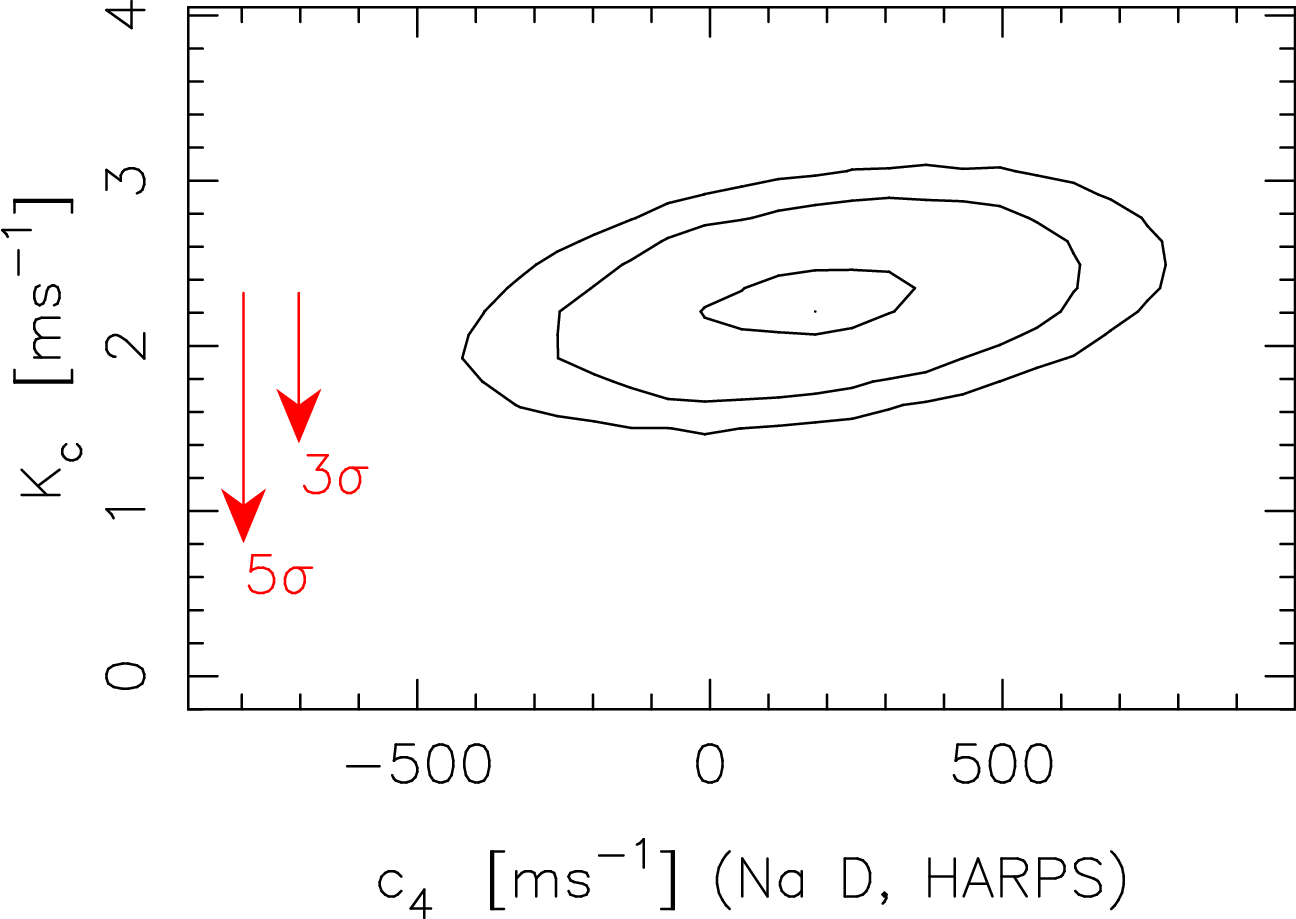}
\includegraphics[angle=270, width=0.195\textwidth]{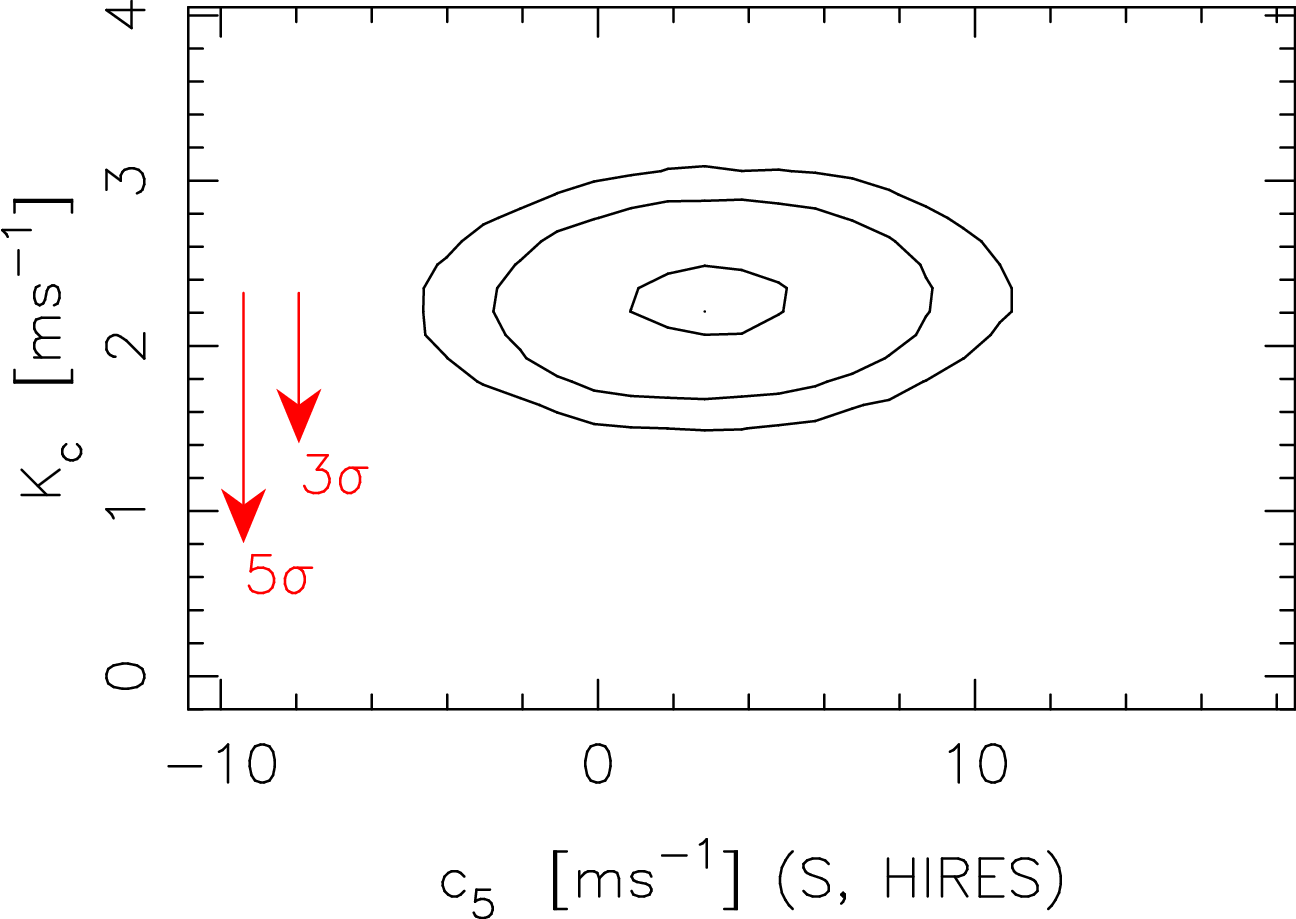}
\includegraphics[angle=270, width=0.195\textwidth]{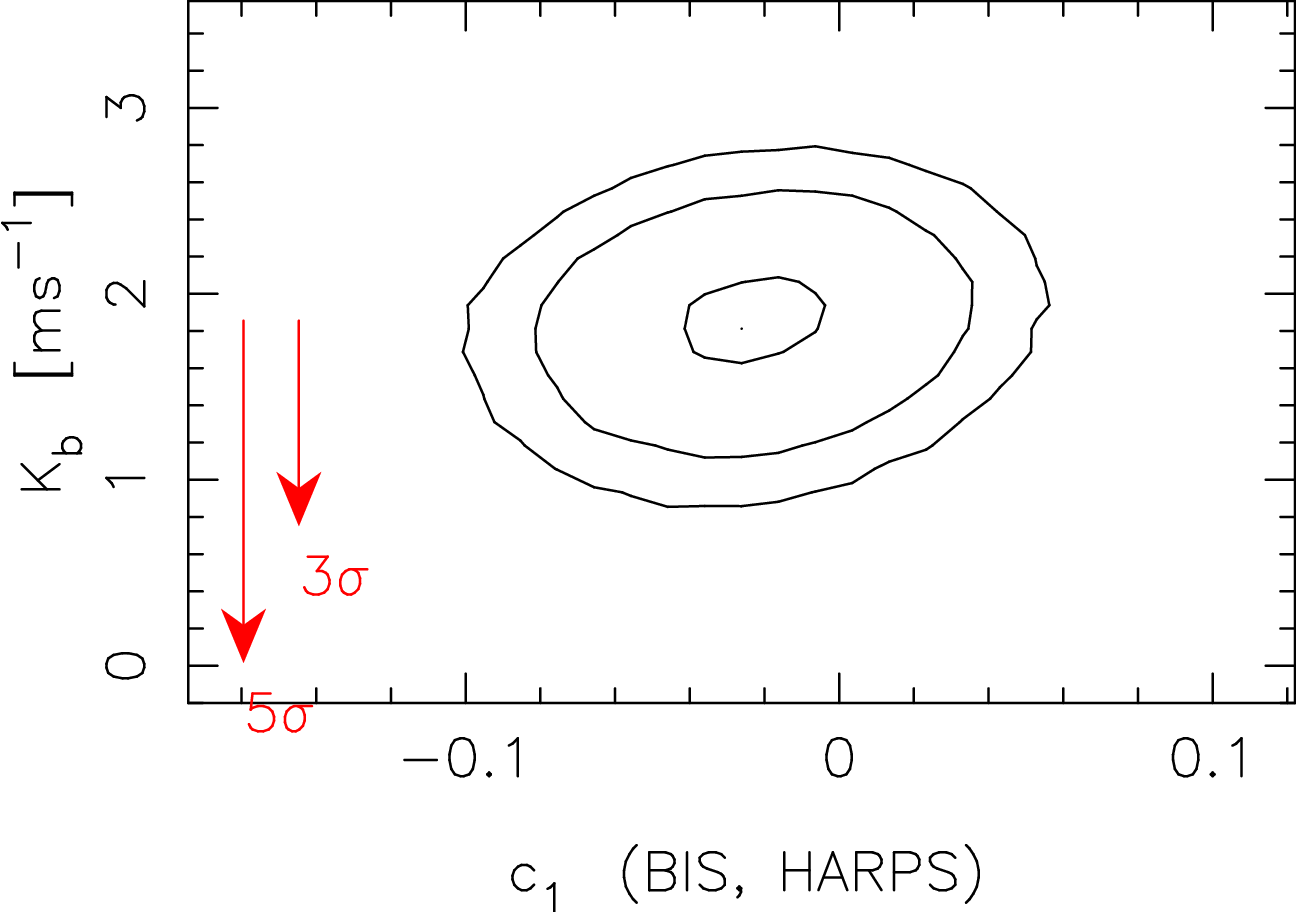}
\includegraphics[angle=270, width=0.195\textwidth]{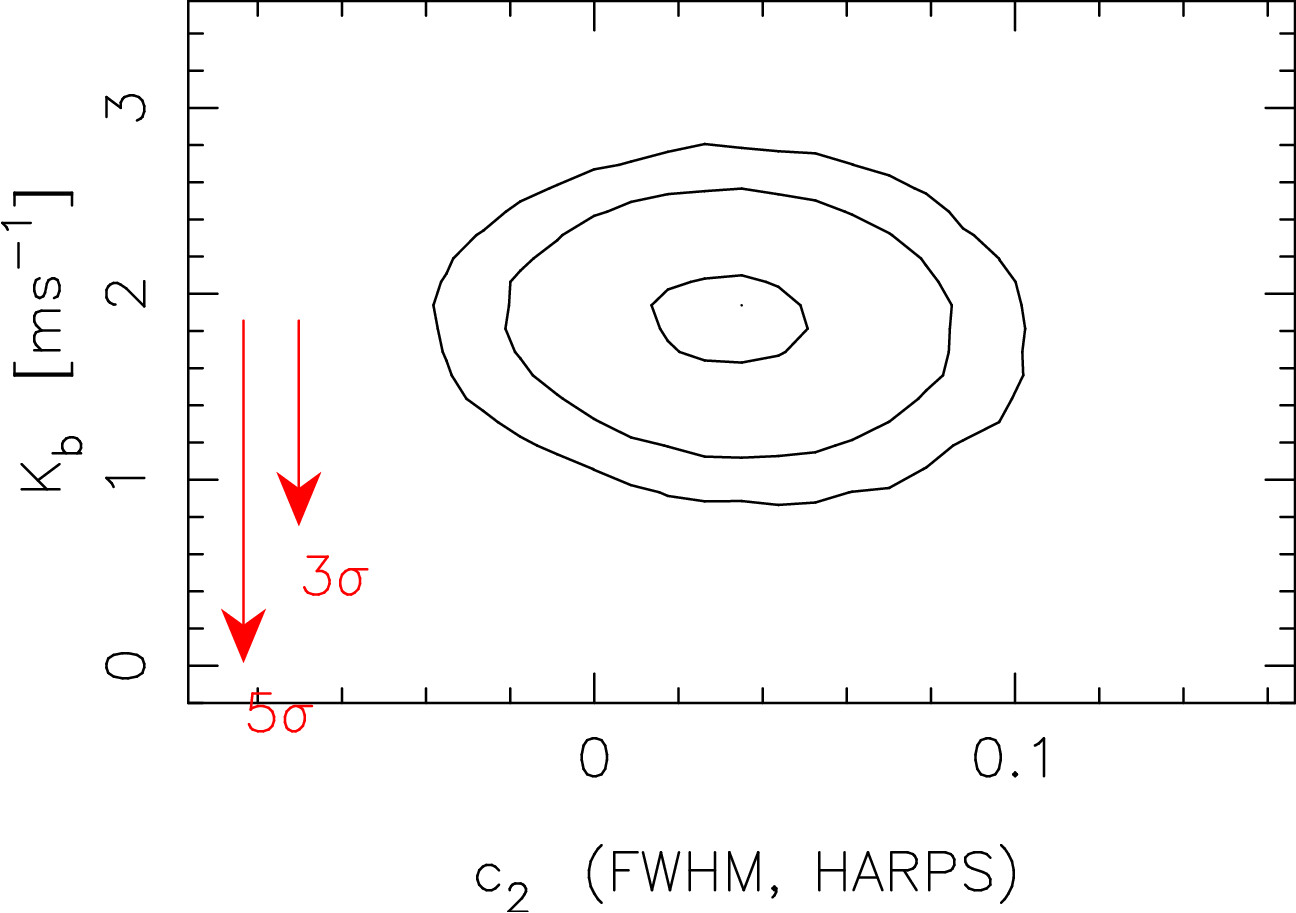}
\includegraphics[angle=270, width=0.195\textwidth]{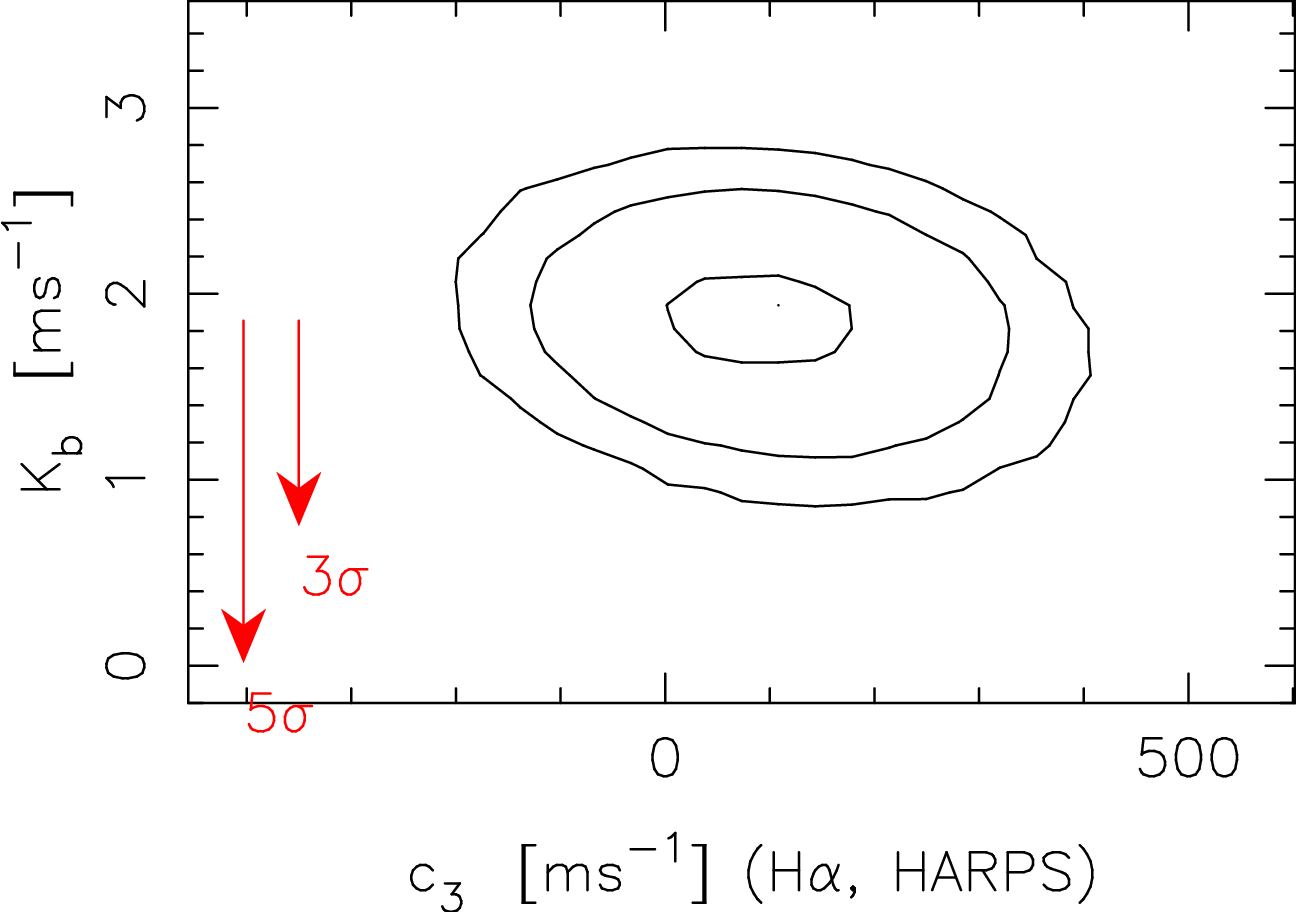}
\includegraphics[angle=270, width=0.195\textwidth]{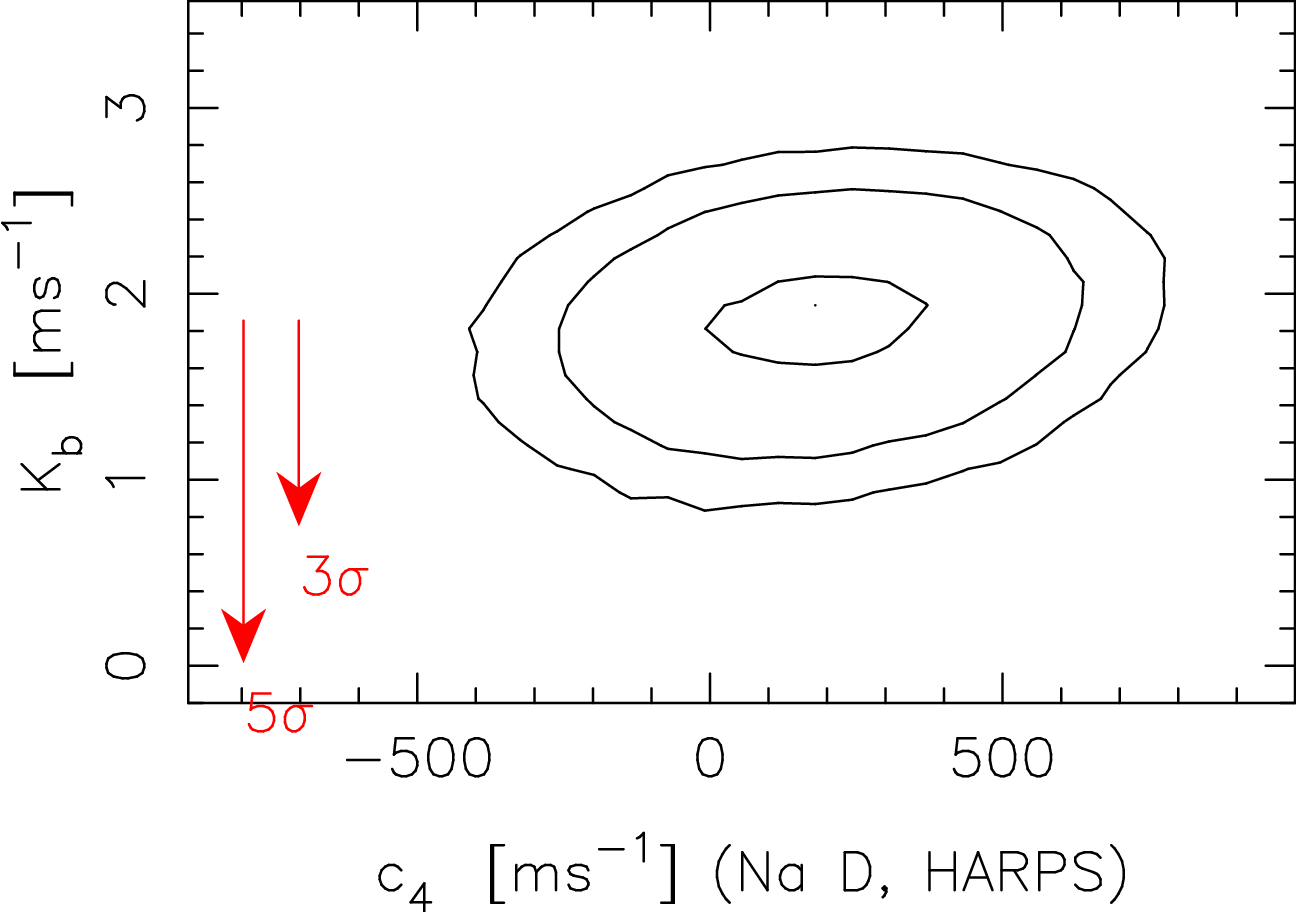}
\includegraphics[angle=270, width=0.195\textwidth]{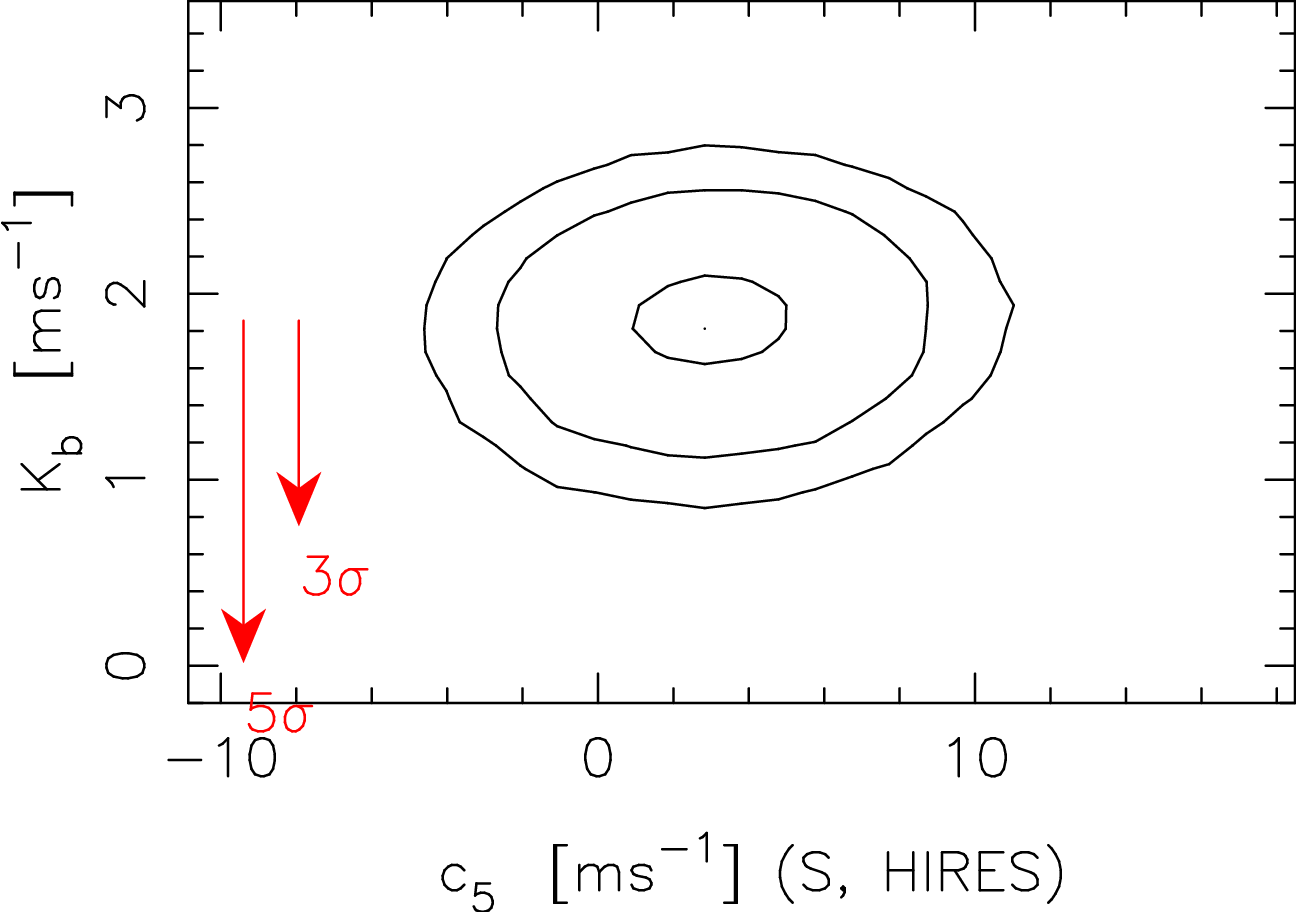}
\caption{Posterior densities and equiprobability contours of the semi-amplitudes
of the planet candidates $K_c$ (top) and $K_b$ (bottom) against the linear
correlation terms defined in the text (x-axis). The contours contain 50\%, 95\%,
and 99\% of the probability density, respectively. The 3$\sigma$ and 5$\sigma$
intervals of the distributions are shown for $K_b$ and $K_c$ to demonstrate how
significantly $K_b$ and $K_c$ differ from $0$.  On the other hand, all $c_i$ are
found to be broadly consistent with $0$.}\label{fig:distributions}
\end{figure*}

In this section we perform a Bayesian analysis to evaluate the significance of
correlations of the RV data with activity indices again assuming the linear
model in Eq.~\ref{eq:model}. As before, we literally use the values provided in
RM15 for simplicity in the discussion. All linear correlation terms ($c_1$
corresponds to HARPS BIS; $c_2$ to HARPS FWHM; $c_3$ to HARPS I$_\alpha$; $c_4$
to HARPS Na D, and $c_5$  to the HARPS S-index) were tested at the same time by
simultaneously including them all as free parameters. As a figure of merit for
model comparison, we obtained the integrated likelihoods of models with and
without signals and linear correlation terms. These integrated likelihoods
(sometimes called \textit{Evidences} $E$) were calculated by setting the priors
as discussed in \citet{tuomi:2013}, and uniform ones for the parameters $c_{i}$.
The  algorithm used for the estimation of the integral is based on a mixture of
Markov Chain Monte Carlo samples from both the  posterior and prior
\citep{newton:raftery:1994}.

Fig.~\ref{fig:distributions} illustrates the posterior
densities of each correlation coefficient c$_i$ against the $K$
semi-amplitudes of the signals at 48.6 (Kapteyn's b) and 120
days (Kapteyn's c). The posterior densities were sampled using
the adaptive-Metropolis posterior sampling algorithm
\citep{haario:2001}. Two features would be expected for a
radial velocity variations signal traced by an activity index.
Firstly, the posterior densities in
Fig.~\ref{fig:distributions} would show a tilted elliptical
shape and the value of the corresponding $c_i$ would be
significantly different from $0$, and secondly, $K$ would be
consistent with $0$ in the sense that 95\% (or 99\%)
equiprobability contours overlapped with zero. Some of the
plots show some mild hints of correlation (tilted ellipses),
but all distributions for the $c_i$ are broadly consistent with
$0$ values. In contrast, the expected value for the
semiamplitudes of Kapteyn b is distinct from $0$ at a $\sim$
5$\sigma$ level (even higher for Kapteyn's c), where $\sigma$
is the standard deviation in of the posterior density in each
$K$ (see Fig~\ref{fig:distributions}). The reason for the
apparent contradiction with the claims in RM15 is explained in
the next section.

Table \ref{tab:evidences} summarizes the model probabilities with linear
correlations and planet signals included.  The evidence ratios between models
with $k$ and $k-1$ signals remain well above any reasonable significance
threshold (eg. model probabilities larger than the 150-1000 factors usually
required to claim a confident detection). The models including linear
correlations (right) have slightly better integrated probabilities than those
without (left), but the improvement is only a factor of $\sim$ 12 when comparing
the models with $k=2$. This negligible level of significance of correlated
variability is again consistent with the confidence level contours of
Fig.~\ref{fig:distributions}, which imply that all $c_i$ are compatible with
$0$.

\begin{deluxetable}{c|cc|cc} 
\tabletypesize{\footnotesize} 
\tablecolumns{5} 
\tablewidth{0pt} 
\tablecaption{Natural logarithms of the integrated model probabilties $E$ and their ratios.
\label{tab:evidences}} 
\tablehead{ Number of Planets
   & \multicolumn{2}{|l|}{Keplerian only} &  \multicolumn{2}{|l}{Keplerian + correlations}\\
%   \hline
k& $\ln E_k$ & $\ln (E_{k}/E_{k-1})$ & $\ln E_k$ & $\ln (E_{k}/E_{k-1})$
}
\startdata 
0 &   -277.7  &      -       &  -273.6 &   -           \\
1 &   -260.1 &   +17.6   & -254.9 &  +18.7      \\
2 &   -238.8 &   +21.3   & -241.3 &  +13.6$^{\dagger}$        
\enddata 
\vspace{-0.8cm} 
\tablecomments{$^\dagger$As a reference, a $\ln (E_{k}/E_{k-1})$ of +13.6 indicates that the
model with $k$ planets has a higher probability than a model with $k-1$ planets 
by a factor $e^{+13.6} = 8.1 \times 10^5$.} 
\end{deluxetable}

%To further verify the status of the Doppler signals in Kapteyn's star, we
%performed our own measurements on activity indices and added 9 new RV
%measurements obtained with the PFS instrument to our previous solution. The
%incorporation of the new PFS data did not improve nor decrease the significance
%of the signals, and the limited number of observations prevented detecting
%further planet candidates. Since the the Keplerian parameters of the planet
%candidates remain very similar to those in \citet{anglada:2014a}, we do not
%include them here for brevity.

\section{Origin of the correlation proposed by RM15}\label{sec:nocorrelations}
\begin{figure}[htb]
\center
\includegraphics[width=0.45\textwidth, clip]{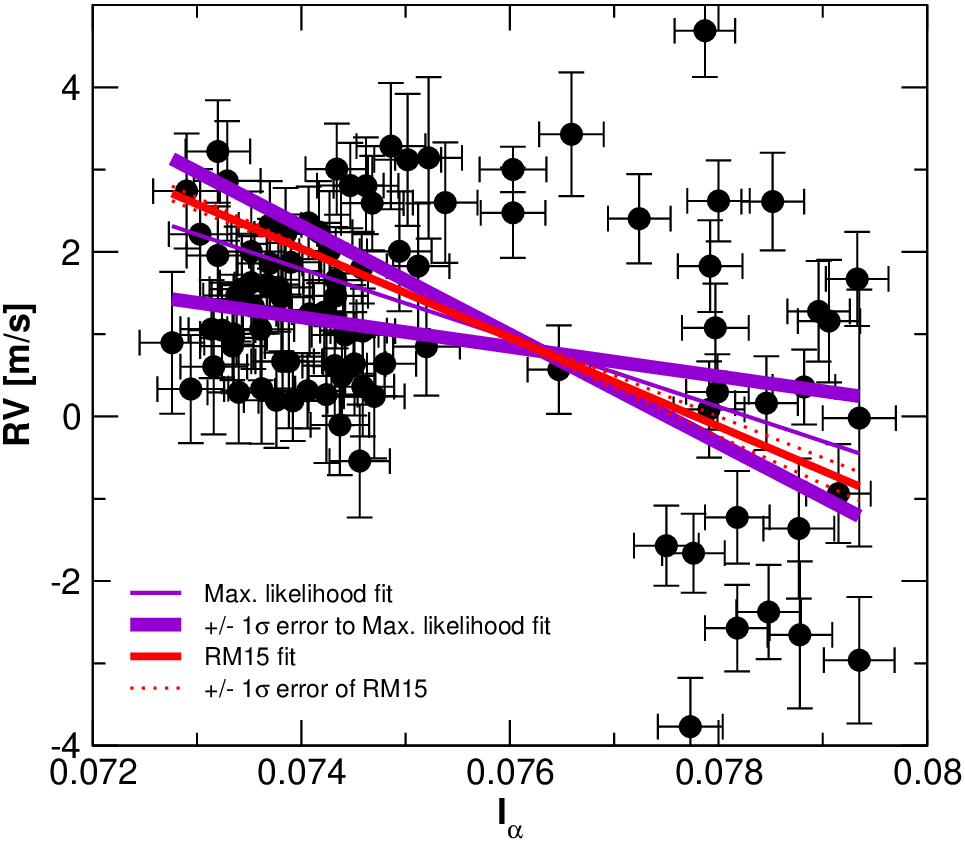}
\caption{Correlation between the I$_\alpha$ index and the
RVs once all signals except Kapteyn's b have been removed
from the data. The thin violet line is the maximum likelihood
fit to the data we obtained, and the thick violet lines represent
alternative fits within 1$\sigma$ values of the obtained
correlation coefficient. The fit proposed by RM15 is represented
by a red line and the 1$\sigma$ representations of their law are
illustrated as dotted red lines. }\label{fig:corplot}
\end{figure}

There is a fundamental difference in the procedure we have used here to assess
the presence of correlations and the one used by RM15. That is, while we used a
global fit to the data to constrain the coefficients, RM15 used the predictions
of the two planet model (with no errors) to perform their analysis.  That is,
RM15's Figure 3 (top-central panel) shows $I_\alpha$ against the Doppler
\textit{model} of planet $b$. In our Figure \ref{fig:corplot}, we show the same 
plot but present the radial velocity measurements after removing all signals
except planet b. The linear correlation law derived from our Bayesian  analysis
in the previous section is presented in violet. Models showing allowed values of
the correlation coefficients at $\pm$ 1$\sigma$ intervals are also represented
as thick violet lines, which visually illustrates the large uncertainty in
those. The best correlation law proposed by RM15 is shown as a red line, and red
dotted lines show values of the coefficient at their reported $\pm$ 1$\sigma$
values. While the linear correlation law reported by RM15 is well within our
1$\sigma$ interval, their reported uncertainties are notoriously underestimated
producing the spurious artifact of significant correlation. This is a direct
consequence of misusing the RV model preductions (no uncertainties), instead of
the actual data on testing the existence of potential correlation laws. We note
for example, that even the Doppler model contains uncertainties, which where
ignored in RM15.

\section{Discussion}\label{sec:conclusions}

We have shown that linear correlations of RVs with activity indicators in the
currently existing data are insignificant for Kapteyn's star's RVs when a global
fit to the data is obtained. This  stands in contrast to the claims made in
RM15, which were based on a number of approximate physical assumptions and the
implementation of \textit{ad hoc} procedures. We also want to stress that
interpretation of the 143d periodicity found by RM15 in several indicators as
rotation period seems premature: alternative periods of 88d, 135d or 270d are
similarly likely, and long-term activity trends cannot be ruled out either. Even
If for the moment we assume that the star rotates at a period of 143d, it is not
straightforward to use this as argument against a Doppler signal close-to
$P_{rot}/3$, because there is no activity signal at $P_{rot}/2$ or $P_{rot}/3$.
Given all these caveats, we consider that the current Doppler data of Kapteyn's
star is most easily explained by the presence of two planets as proposed in
\citet{anglada:2014a} rather than activity induced variability as proposed by
RM15.

A clear distinction must be made between the statistical significance of RV
signals and the physical presence of planets (together with the merit of their
detection or falsification). We advocate for comprehensive scientific
discussions about the former instead of running into premature and unsupported
statements about the latter. We conclude by emphasizing that the intention of
this paper is not to rescue the planetary status of Kapteyn's b or any other
planet detection, but to stress the importance of objective global analysis
techniques in serious scientific discussions.

\bibliographystyle{apj}

\end{document}